\documentclass[conference]{IEEEtran}%
\pdfoutput=1
\usepackage{amsfonts}
\usepackage{amsmath}
\usepackage{amssymb}
\usepackage{graphicx}%
\providecommand{\U}[1]{\protect\rule{.1in}{.1in}}

\begin{document}

\title{The free space optical interference channel}
\author{\IEEEauthorblockN{Saikat Guha}
\IEEEauthorblockA{Disruptive Information Processing Technologies Group\\
Raytheon BBN Technologies\\
Cambridge, Massachusetts, USA\\
Email: sguha@bbn.com}
\and
\IEEEauthorblockN{Ivan Savov and Mark M. Wilde}
\IEEEauthorblockA{School of Computer Science\\
McGill University\\
Montreal, Qu\'{e}bec, Canada\\
Emails: ivan.savov@mcgill.ca; mwilde@cs.mcgill.ca}}
\maketitle

\begin{abstract}
Semiclassical models for multiple-user optical communication cannot assess the
ultimate limits on reliable communication as permitted by the laws of physics.
In all optical communications settings that have been analyzed within a
quantum framework so far, the gaps between the quantum limit to the capacity
and the Shannon limit for structured receivers become most significant in
the low photon-number regime. Here, we present a quantum treatment of a
multiple-transmitter multiple-receiver multi-spatial-mode free-space
interference channel with diffraction-limited loss and a thermal background.
We consider the performance of a laser-light (coherent state) encoding in
conjunction with various detection strategies such as homodyne, heterodyne,
and joint detection. Joint detection outperforms both homodyne and heterodyne
detection whenever the channel exhibits \textquotedblleft very
strong\textquotedblright\ interference. We determine the capacity region for
homodyne or heterodyne detection when the channel has \textquotedblleft
strong\textquotedblright\ interference,\ and we conjecture the existence of a
joint detection strategy that outperforms the former two strategies in this
case. Finally, we determine the Han-Kobayashi achievable rate regions for both
homodyne and heterodyne detection and compare them to a region achievable by a
conjectured joint detection strategy. In these latter cases, we determine
achievable rate regions if the receivers employ a recently discovered
min-entropy quantum simultaneous decoder.

\end{abstract}

\section{Introduction}

The principal goals of information theory are to determine the ultimate limits
on reliable communication and to find ways of approaching these limits in
practice. Point-to-point optical communication using laser-light modulation in
conjunction with direct-detection and coherent-detection receivers has been
studied in detail using the semiclassical theory of photodetection~\cite{GK95}%
. This approach treats light as a classical electromagnetic field, and the
fundamental noise encountered in photodetection is the shot noise associated
with the discreteness of the electron charge.

These semiclassical treatments for systems that exploit
classical-light modulation and conventional receivers (direct, homodyne, or
heterodyne) have had some success,
but we should recall that electromagnetic waves are quantized, and
the correct assessment of systems that use non-classical light sources
and/or general optical measurements requires a full quantum-mechanical
framework~\cite{S09}. Consider several recent theoretical studies on the
point-to-point~\cite{GGLMSY04, G10}, broadcast~\cite{GSE07} and
multiple-access~\cite{Y05} bosonic channels with linear loss and a thermal
background. These studies have shown that achievable communication rates
surpass what can be obtained with conventional receivers.
Prior work has established that Holevo information rates are achievable for
information transmission on general quantum point-to-point~\cite{SW97,H98},
broadcast~\cite{YHD06} and multiple-access~\cite{YHD08} channels. In each
case, the Holevo information rates are an upper bound to the Shannon rates
computed for any specific transmitter-modulation receiver-measurement pair.
For the general quantum channel, attaining Holevo information rates may
require collective measurements (a joint detection) across the channel outputs.

The next level of complexity beyond the point-to-point, broadcast, or
multiple-access channels is arguably captured by the interference
channel~\cite{C75,S81,HK81}. This channel, in general, can have $M$ senders and $M$ receivers
($M\geq2$), where each sender would like to communicate only with a partner
receiver,\footnote{Note that this interference channel setting has been
extensively analyzed in network information theory, for scaling behavior of
the total source-to-destination capacity and communication latency as $M$
grows, for randomly distributed $M$ source-destination node pairs in a given
network area.} but most research focuses on the special case of
two senders and two receivers. The Gaussian interference channel has been analyzed in depth in
the classical information-theory literature, and this model readily applies
for an optical interference channel with coherent-state inputs and coherent
detection. Calculating the capacity of this channel in the general
case has been an open problem for some time, but several researchers have
found it in the special cases of \textquotedblleft very
strong\textquotedblright~\cite{C75} and \textquotedblleft
strong\textquotedblright\ interference~\cite{S81,HK81}. Also, Han and
Kobayashi determined the best inner bound on the channel's capacity, by
having each receiver partially decode the message of the other sender along
with a full decoding of the message of the partner sender~\cite{HK81}.
However, all of these strategies assume that the underlying channel is
classical,
and we would expect a quantum strategy with a power-constrained encoding and
collective measurement at the receivers to outperform such strategies.

In this paper, we consider a pure-loss thermal-noise bosonic interference
channel, particularly in the context of free-space (wireless) terrestrial
optical communications. We assume a coherent-state encoding throughout this
paper. We find achievable rate regions for the two-sender two-receiver channel
with coherent-detection receivers, for the \textquotedblleft very
strong\textquotedblright\ and \textquotedblleft strong\textquotedblright%
\ interference regimes. We find the rate region achievable with a
joint-detection receiver (JDR) under \textquotedblleft very
strong\textquotedblright\ interference. We also determine a \textquotedblleft
min-entropy\textquotedblright\ JDR-achievable rate region for the
\textquotedblleft strong\textquotedblright\ interference regime by exploiting
a recent result from Ref.~\cite{FHSSW11}, and we find a conjectured
JDR-achievable region were a conjecture from Ref.~\cite{FHSSW11} regarding
quantum simultaneous decoding true (c.f, page 4-15 of
Ref.~\cite{el2010lecture}\ for a classical simultaneous decoder). Next, we
evaluate the Han-Kobayashi rate region for homodyne and heterodyne detection
in the general case, and we show an achievable rate region using a
\textquotedblleft min-entropy\textquotedblright\ quantum simultaneous decoder
from Ref.~\cite{FHSSW11}. Finally, we conjecture a Han-Kobayashi-like
achievable rate region using Conjecture~2 from Ref.~\cite{FHSSW11}.
Our results here differ from those in Ref.~\cite{FHSSW11}---there some of us
considered the general quantum interference channel whereas here we
consider specifically a bosonic interference channel.

\section{A Free-Space Optical Interference Channel}

Consider a range-$L$ line-of-sight free-space optical channel with hard
circular transmit and receive apertures of areas $A_{t}$ and $A_{r}$
respectively. Assume $\lambda$-center-wavelength quasi-monochromatic
transmission. In the near-field propagation regime (Fresnel number product,
$D_{f}\equiv A_{t}A_{r}/(\lambda{L})^{2}\gg1$), a normal-mode decomposition of
the free-space optical channel yields $M\approx2D_{f}$ orthogonal spatio-polarization
transmitter-to-receiver modes ($D_{f}$ spatial modes, each of two orthogonal
polarizations) with near-unity transmitter-to-receiver power transmissivities
($\eta_{m}\approx1$). In the far-field propagation regime ($D_{f}\ll1$), only
two orthogonal spatial modes (one of each orthogonal polarization) have
appreciable power transmissivity ($\eta\approx D_{f}$ for each mode).

Sender $m$ modulates her information on the $m^{\mathrm{th}}$
transmitter-pupil spatial mode, and Receiver $m$ separates and demodulates
information from the corresponding receiver-pupil spatial mode. With perfect
spatial-mode control at the transmitter and perfect mode separation at the
receiver, the orthogonal spatial modes can be thought of as independent
parallel channels with no cross talk. However, imperfect (slightly
non-orthogonal) mode generation or imperfect mode separation can result in
cross talk (interference) between the $M$ channels.

We take our interference channel model as a passive linear mixing of the input
modes along with the possibility of a thermal environment adding
zero-mean, isotropic Gaussian noise.
Although the results here apply to cyclic interference channels with
$M$ senders and $M$ receivers,
for simplicity, we limit ourselves to the $M=2$ case, in which case the
channel model reduces to
\begin{align}
\hat{b}_{1} &  =\sqrt{\eta_{11}}\hat{a}_{1}+\sqrt{\eta_{21}}\hat{a}_{2}%
+\sqrt{\bar{\eta}_{1}}\hat{\nu}_{1},\\
\hat{b}_{2} &  =\sqrt{\eta_{12}}\hat{a}_{1}-\sqrt{\eta_{22}}\hat{a}_{2}%
+\sqrt{\bar{\eta}_{2}}\hat{\nu}_{2},
\end{align}
where $\eta_{11}, \eta_{12}, \eta_{21}, \eta_{22}, \bar{\eta}_{1}, \bar{\eta}_{2} \in \mathbb{R}^+$,
$\sqrt{\eta_{11} \eta_{12}} = \sqrt{\eta_{21} \eta_{22}}$,
$
\bar{\eta}_{1}\equiv1-\eta_{11}-\eta_{21}$, and $\bar{\eta}_{2}%
\equiv1-\eta_{12}-\eta_{22}$.
The following conditions ensure that the network is passive:%
\[
\eta_{11}+\eta_{12}\leq1,\ \ \ \eta_{11}+\eta_{21}\leq1,\ \ \ \eta_{22}%
+\eta_{21}\leq1,\ \ \ \eta_{22}+\eta_{12}\leq1.
\]
We constrain the mean photon number of the
transmitters\ $\hat{a}_{1}$ and $\hat{a}_{2}$ to be $N_{S_{1}}$ and $N_{S_{2}%
}$ photons per mode, respectively, the environment modes ${\hat{\nu}}_{1}$ and
${\hat{\nu}}_{2}$ are in statistically independent zero-mean thermal states
\cite{S09} with respective mean photon numbers $N_{B_{1}}$ and $N_{B_{2}}$ per
mode.

Note that for a coherent-state encoding and coherent-detection at both
receivers, the above model is a special case of a complex Gaussian
interference channel, and we can study its capacity regions in various
settings by applying the results from Refs.~\cite{C75,S81,HK81}. If the
senders prepare their inputs in coherent states $\left\vert \alpha
_{1}\right\rangle $ and $\left\vert \alpha_{2}\right\rangle $, with
$\alpha_{1},\alpha_{2}\in{\mathbb{R}}$, and both receivers perform
real-quadrature homodyne detection on their respective modes, the result is a
classical Gaussian interference channel~\cite{S09}, where Receivers~1 and 2
obtain respective conditional Gaussian random variables $Y_{1}$ and $Y_{2}$
distributed as
\begin{align*}
Y_{1} &  \sim\mathcal{N}\left(  \sqrt{\eta_{11}}\alpha_{1}+\sqrt{\eta_{21}%
}\alpha_{2},\ \left(  2\bar{\eta}_{1}N_{B_{1}}+1\right) / 4 \right)
,\\
Y_{2} &  \sim\mathcal{N}\left(  \sqrt{\eta_{12}}\alpha_{2}+\sqrt{\eta_{22}%
}\alpha_{1},\ \left(  2\bar{\eta}_{2}N_{B_{2}}+1\right) / 4  \right)  ,
\end{align*}
where the \textquotedblleft$+1$\textquotedblright\ term in the noise variances
arises physically from the zero-point fluctuations of the vacuum. Suppose that
the senders again encode their signals as coherent states $\left\vert
\alpha_{1}\right\rangle $ and $\left\vert \alpha_{2}\right\rangle $, but this
time with $\alpha_{1},\alpha_{2}\in\mathbb{C}$, and that the receivers both
perform heterodyne detection. This results in a classical complex Gaussian
interference channel~\cite{S09}, where Receivers~1 and 2 detect respective
conditional complex Gaussian random variables $Z_{1}$ and $Z_{2}$, whose real
parts are distributed as
\[
\operatorname{Re}\left\{  Z_{m}\right\}  \sim\mathcal{N}\left(  \mu
_{m},\ \left(  \bar{\eta}_{m}N_{B_{m}}+1\right) \!/  2  \right)  ,
\]
where $m\in\left\{  1,2\right\}  $, $\mu_{1}\equiv\sqrt{\eta_{11}%
}\operatorname{Re}\left\{  \alpha_{1}\right\}  +\sqrt{\eta_{21}}%
\operatorname{Re}\left\{  \alpha_{2}\right\}  $, $\mu_{2}\equiv\sqrt{\eta
_{12}}\operatorname{Re}\left\{  \alpha_{1}\right\}  +\sqrt{\eta_{22}%
}\operatorname{Re}\left\{  \alpha_{2}\right\}  $, and the imaginary parts of
$Z_{1}$ and $Z_{2}$ are distributed with the same variance as their real
parts, and their respective means are $\sqrt{\eta_{11}}\operatorname{Im}%
\left\{  \alpha_{1}\right\}  +\sqrt{\eta_{21}}\operatorname{Im}\left\{
\alpha_{2}\right\}  $ and $\sqrt{\eta_{12}}\operatorname{Im}\left\{
\alpha_{1}\right\}  +\sqrt{\eta_{22}}\operatorname{Im}\left\{  \alpha
_{2}\right\}  $. The factor of 1/2 in the noise variances is due to the
attempt to measure both quadratures of the field simultaneously~\cite{S09}.

\section{Very Strong Interference}

\label{sec:very-strong-int}Carleial determined the capacity region of a
classical Gaussian interference channel in the case of \textquotedblleft very
strong\textquotedblright\ interference \cite{C75}. Suppose that $X_{1}$ and
$X_{2}$ are the input random variables for Senders~1 and 2 and that $B_{1}$
and $B_{2}$ represent the outputs for Receivers~1 and 2, respectively. Then
the most general way to state the condition for very strong interference is
that the following information inequalities should hold for all input
distributions $p_{X_{1}}\!\left(  x_{1}\right)  $ and $p_{X_{2}}\!\left(
x_{2}\right)  $~\cite{el2010lecture}:%
\begin{align}
I\left(  X_{1};B_{1}|X_{2}\right)   &  \leq I\left(  X_{1};B_{2}\right)
,\label{eq:VSI-1}\\
I\left(  X_{2};B_{2}|X_{1}\right)   &  \leq I\left(  X_{2};B_{1}\right)  .
\label{eq:VSI-2}%
\end{align}
Carleial proved the surprising result \cite{C75} that the capacity region of
the interference channel under this setting is the convex closure of positive
rate pairs $\left(  R_{1},R_{2}\right)  $ such that%
\begin{equation}
R_{1}\leq I\left(  X_{1};B_{1}|X_{2}\right)  ,\ \ \ \ R_{2}\leq I\left(
X_{2};B_{2}|X_{1}\right)  , \label{eq:carleial-rate-1}%
\end{equation}
for some input distributions $p_{X_{1}}\!\left(  x_{1}\right)  $ and $p_{X_{2}%
}\!\left(  x_{2}\right)  $. The coding strategy is simply for each receiver to
decode first what the other sender is transmitting, remove this signal, and
then decode the message intended for him.

The conditions in (\ref{eq:VSI-1}-\ref{eq:VSI-2}) translate to the following
ones for the case of coherent-state encoding and coherent detection:%
\begin{align*}
\frac{\eta_{21}}{\eta_{22}}  &  \geq\frac{4^{i}\eta_{11}N_{S_{1}}+2^{i}%
\bar{\eta}_{1}N_{B_{1}}+1}{2^{i}\bar{\eta}_{2}N_{B_{2}}+1},\\
\frac{\eta_{12}}{\eta_{11}}  &  \geq\frac{4^{i}\eta_{22}N_{S_{2}}+2^{i}%
\bar{\eta}_{2}N_{B_{2}}+1}{2^{i}\bar{\eta}_{1}N_{B_{1}}+1},
\end{align*}
and the capacity region becomes%
\begin{align}
R_{1}  &  \leq\frac{1}{2^{i}}\ln\left(  1+\frac{4^{i}\eta_{11}N_{S_{1}}}%
{2^{i}\bar{\eta}_{1}N_{B_{1}}+1}\right)  ,\label{eq:hom-VSI-cap-1}\\
R_{2}  &  \leq\frac{1}{2^{i}}\ln\left(  1+\frac{4^{i}\eta_{22}N_{S_{2}}}%
{2^{i}\bar{\eta}_{2}N_{B_{2}}+1}\right)  , \label{eq:hom-VSI-cap-2}%
\end{align}
where $i=1$ for homodyne detection and $i=0$ for heterodyne detection.

\begin{figure}[ptb]
\begin{center}
\includegraphics[
natheight=2.753600in,
natwidth=5.813300in,
height=1.644in,
width=3.4411in
]{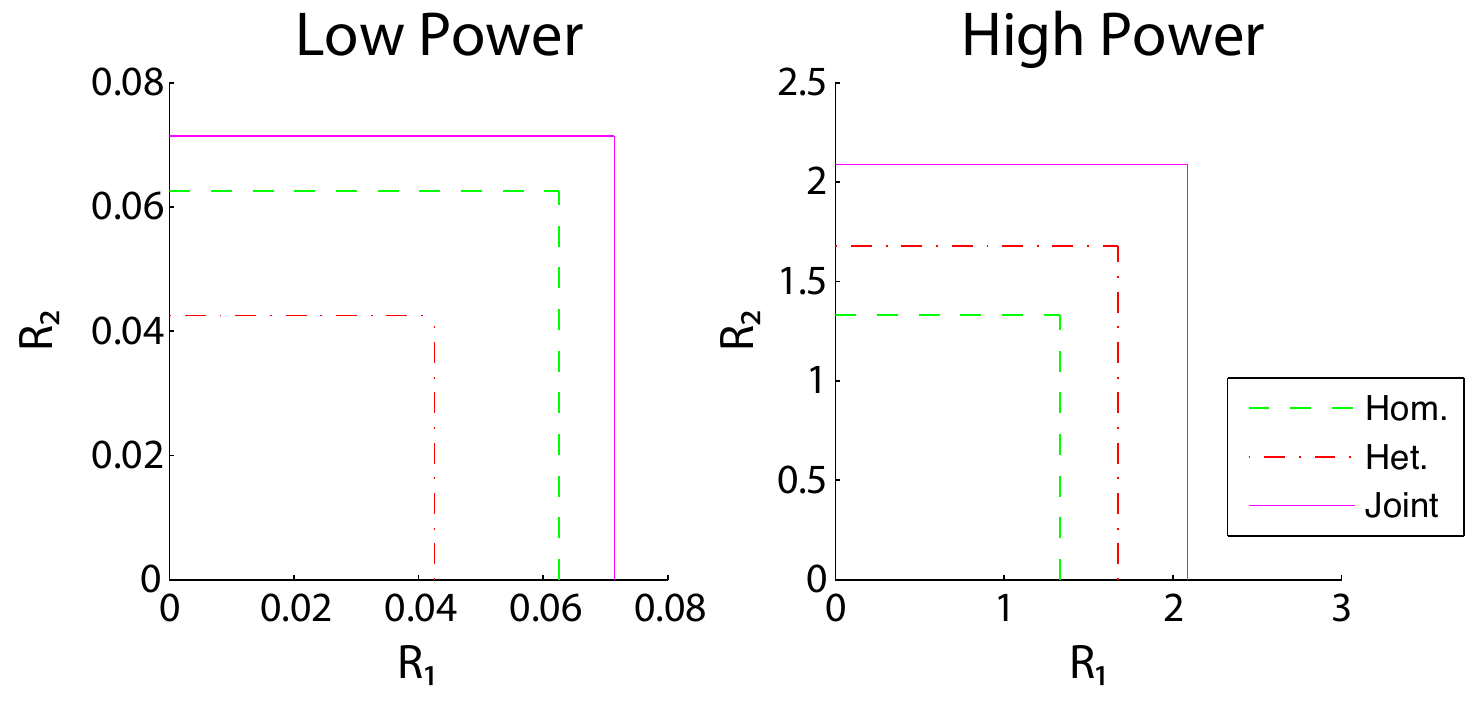}
\end{center}
\caption{Capacity regions for coherent-state encodings and coherent detection,
and achievable rate regions for coherent-state encodings and joint-detection
receivers---both with $\eta_{11}=\eta_{22}= 1 / 16$ and $\eta_{12}%
=\eta_{21}=1/2$ (\textquotedblleft very strong\textquotedblright%
\ interference for coherent detection and such that (\ref{eq:VSI-g-1}%
-\ref{eq:VSI-g-2}) hold). The LHS\ displays these regions in a low-power
regime with $N_{S_{1}}=N_{S_{2}}=1$ and $N_{B_{1}}=N_{B_{2}}=1$, and the
RHS\ displays these regions in a high-power regime where $N_{S_{1}}=N_{S_{2}%
}=100$. Homodyne detection outperforms heterodyne detection in the low-power
regime because it has a reduced detection noise, while heterodyne detection
outperforms homodyne detection in the high-power regime because its has an
increased bandwidth.}%
\label{fig:carleial}%
\end{figure}

We can also consider the case when the senders employ coherent-state encodings
and the receivers employ a joint-detection strategy on all of their respective
channel outputs. The conditions in (\ref{eq:VSI-1}-\ref{eq:VSI-2}) readily
translate to this quantum setting where we now consider $B_{1}$ and $B_{2}$ to
be quantum systems, and the information quantities in (\ref{eq:VSI-1}%
-\ref{eq:VSI-2}) and (\ref{eq:carleial-rate-1}) now become Holevo
informations~\cite{FHSSW11}. Theorem~6\ in Ref.~\cite{FHSSW11}\ provides a
simple proof that Carleial's result applies in the quantum domain to these
Holevo informations. So, the conditions in (\ref{eq:VSI-1}-\ref{eq:VSI-2})
when restricting to coherent-state encodings translate to%
\begin{align}
&  g\left(  \eta_{22}N_{S_{2}}+\bar{\eta}_{2}N_{B_{2}}\right)  -g\left(
\bar{\eta}_{2}N_{B_{2}}\right)  \label{eq:VSI-g-1}\\
&  \leq g\left(  \eta_{21}N_{S_{2}}+\eta_{11}N_{S_{1}}+\bar{\eta}_{1}N_{B_{1}%
}\right)  -g\left(  \eta_{11}N_{S_{1}}+\bar{\eta}_{1}N_{B_{1}}\right)
,\nonumber\\
&  g\left(  \eta_{11}N_{S_{1}}+\bar{\eta}_{1}N_{B_{1}}\right)  -g\left(
\bar{\eta}_{1}N_{B_{1}}\right)  \label{eq:VSI-g-2}\\
&  \leq g\left(  \eta_{12}N_{S_{1}}+\eta_{22}N_{S_{2}}+\bar{\eta}_{2}N_{B_{2}%
}\right)  -g\left(  \eta_{22}N_{S_{2}}+\bar{\eta}_{2}N_{B_{2}}\right)
.\nonumber
\end{align}
where $g\left(  N\right)$ is the entropy of a thermal state with mean photon number $N$:
$$
g\left(  N\right)  \equiv\left(  N+1\right)  \ln\left(
N+1\right)  -N\ln\left(  N\right).$$
An achievable rate region is then%
\begin{align*}
R_{1} &  \leq g\left(  \eta_{11}N_{S_{1}}+\bar{\eta}_{1}N_{B_{1}}\right)
-g\left(  \bar{\eta}_{1}N_{B_{1}}\right)  ,\\
R_{2} &  \leq g\left(  \eta_{22}N_{S_{2}}+\bar{\eta}_{2}N_{B_{2}}\right)
-g\left(  \bar{\eta}_{2}N_{B_{2}}\right)  .
\end{align*}
These rates are
achievable using a coherent-state encoding, but not necessarily optimal
(though they would be optimal if the minimum-output entropy conjecture from
Refs.~\cite{GGLMSY04,GHLM10}\ were true). Nevertheless, these rates always
beat the rates from homodyne and heterodyne detection, and
Figure~\ref{fig:carleial}\ displays examples of the capacity (and achievable
rate) regions in the low- and high-power regimes.

\section{Strong Interference}

\label{sec:strong-int}Sato~\cite{S81} and Han-Kobayashi~\cite{HK81}
independently determined the capacity of a classical Gaussian interference
channel under \textquotedblleft strong\textquotedblright\ interference. A
channel has \textquotedblleft strong\textquotedblright\ interference if the
following information inequalities hold for all $p_{X_{1}}\!\left(
x_{1}\right)  $ and $p_{X_{2}}\!\left(  x_{2}\right)  $~\cite{el2010lecture}:%
\begin{align}
I\left(  X_{1};B_{1}|X_{2}\right)   &  \leq I\left(  X_{1};B_{2}|X_{2}\right)
,\label{eq:SI-1}\\
I\left(  X_{2};B_{2}|X_{1}\right)   &  \leq I\left(  X_{2};B_{1}|X_{1}\right)
.\label{eq:SI-2}%
\end{align}
The capacity region of the classical interference channel under this setting
is the convex closure of positive rate pairs $\left(  R_{1},R_{2}\right)  $
such that \cite{S81,HK81,el2010lecture}:%
\begin{gather}
R_{1}\leq I\left(  X_{1};B_{1}|X_{2}\right)  ,\ \ \ R_{2}\leq I\left(
X_{2};B_{2}|X_{1}\right)  ,\label{eq:strong-rate-1}\\
R_{1}+R_{2}\leq\min\left\{  I\left(  X_{1}X_{2};B_{1}\right)  ,I\left(
X_{1}X_{2};B_{2}\right)  \right\}  .\label{eq:strong-rate-3}%
\end{gather}

The conditions in (\ref{eq:SI-1}-\ref{eq:SI-2}) translate to the following
ones for coherent-state encoding and coherent detection:%
\[
\frac{\eta_{21}}{\eta_{22}}\geq\frac{2^{i}\bar{\eta}_{1}N_{B_{1}}+1}{2^{i}%
\bar{\eta}_{2}N_{B_{2}}+1},\ \ \ \ \ \ \ \ \frac{\eta_{12}}{\eta_{11}}%
\geq\frac{2^{i}\bar{\eta}_{2}N_{B_{2}}+1}{2^{i}\bar{\eta}_{1}N_{B_{1}}+1},
\]
and the capacity region has the two inequalities in (\ref{eq:hom-VSI-cap-1}%
-\ref{eq:hom-VSI-cap-2}) and an additional bound on the sum rate:%
\[
R_{1}+R_{2}\leq\frac{1}{2^{i}}\min\left\{
\begin{array}
[c]{c}%
\ln\left(  1+4^{i}\frac{\eta_{11}N_{S_{1}}+\eta_{21}N_{S_{2}}}{2^{i}\bar{\eta
}_{1}N_{B_{1}}+1}\right)  ,\\
\ \ \ \ln\left(  1+4^{i}\frac{\eta_{22}N_{S_{2}}+\eta_{12}N_{S_{1}}}{2^{i}%
\bar{\eta}_{2}N_{B_{2}}+1}\right)
\end{array}
\right\}  ,
\]
where again $i=1$ for homodyne detection and $i=0$ for heterodyne detection.

\begin{figure}[ptb]
\begin{center}
\includegraphics[
natheight=4.452900in,
natwidth=10.773800in,
width=3.4904in
]{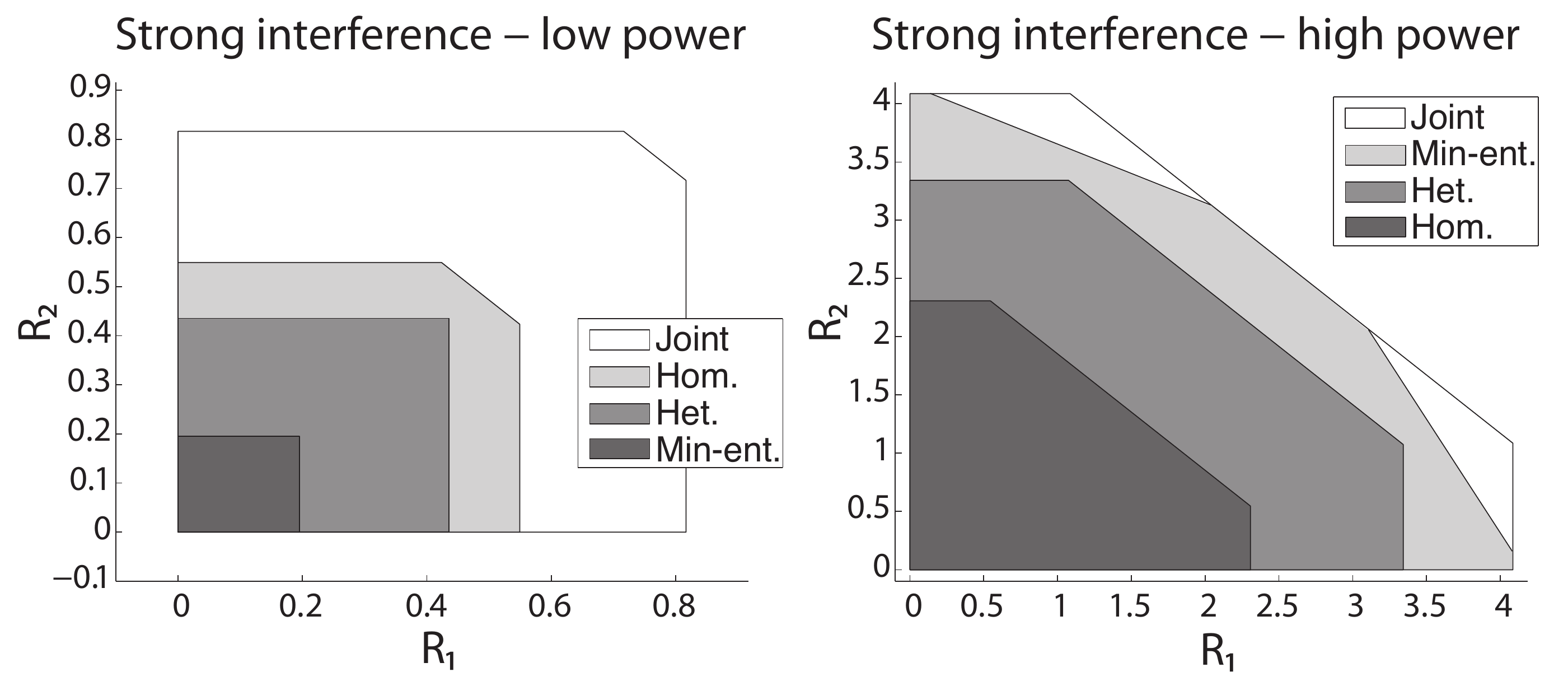}
\end{center}
\caption{The parameters of the free-space channel are
$N_{B_{1}}=N_{B_{2}}=1$, $\eta_{11}=\eta_{22}=0.3$, $\eta
_{21}=\eta_{12}=0.6$, $N_{S_{1}}=N_{S_{2}%
}=2$ for the low-power regime, and $N_{S_{1}}=N_{S_{2}%
}=100$ for the high-power regime. The LHS\ figure depicts the \textquotedblleft
strong\textquotedblright\ interference capacity regions  in the low-power regime for homodyne and
heterodyne detection, the achievable rate region with the conjectured joint
detection, and the convex hull over regions arising from different min-entropy
decoding strategies. Homodyne detection outperforms
heterodyne detection in this regime, and the min-entropy decoders perform the
worst. The RHS\ figure displays the same regions in the high-power regime.
Heterodyne detection outperforms homodyne detection, and interestingly, the
min-entropy decoders outperform heterodyne detection.}%
\label{fig:strong-int}%
\end{figure}

The situation for a joint-detection strategy over all of the channel outputs
becomes more complicated for the case of \textquotedblleft
strong\textquotedblright\ interference, because we require a quantum
simultaneous decoder~\cite{FHSSW11} in order to achieve the information rates
in (\ref{eq:strong-rate-1}-\ref{eq:strong-rate-3}) with $B_{1}$ and $B_{2}$
becoming quantum systems. Such a simultaneous decoder is analogous to a
classical simultaneous decoder (e.g., see page 4-15 of
Ref.~\cite{el2010lecture}), but we have not yet been able to prove the
existence of it in the quantum case (see Conjecture~2 of Ref.~\cite{FHSSW11}).
Yet, we do have an achievable simultaneous decoding strategy expressed in
terms of min-entropies (see Theorem~4 of Ref.~\cite{FHSSW11}), where the
min-entropy of a probability distribution is the negative logarithm of the
probability of its mode \cite{R60}, and, as a simple extension of this idea,
the min-entropy of a density operator is the negative logarithm of its maximum
eigenvalue. For a thermal state with average photon number $N_{B}$, its
min-entropy is $\ln\left(  N_{B}+1\right)  $, and this result allows us to
determine an achievable rate region with a simultaneous decoding strategy
(similar to the simultaneous decoding inner bound on page 6-7 of
Ref.~\cite{el2010lecture}):%
\begin{align*}
R_{1} &  \leq\ln\left(  \eta_{11}N_{S_{1}}+\bar{\eta}_{1}N_{B_{1}}+1\right)
-g\left(  \bar{\eta}_{1}N_{B_{1}}\right)  ,\\
R_{2} &  \leq\ln\left(  \eta_{22}N_{S_{2}}+\bar{\eta}_{2}N_{B_{2}}+1\right)
-g\left(  \bar{\eta}_{2}N_{B_{2}}\right),
\end{align*}
\begin{multline*}
R_{1}+R_{2}\leq\\
\min\left\{
\begin{array}
[c]{c}%
g\left(  \eta_{11}N_{S_{1}}+\eta_{21}N_{S_{2}}+\bar{\eta}_{1}N_{B_{1}}\right)
-g\left(  \bar{\eta}_{1}N_{B_{1}}\right)  ,\\
g\left(  \eta_{22}N_{S_{2}}+\eta_{12}N_{S_{1}}+\bar{\eta}_{2}N_{B_{2}}\right)
-g\left(  \bar{\eta}_{2}N_{B_{2}}\right)
\end{array}
\right\}  .
\end{multline*}
This is one particular variation of an achievable strategy in which we have
the bounds on the individual rates expressed in terms of min-entropies and the
bound on the sum rate expressed with von Neumann entropies, though note that
there are other variations we could consider in light of Theorem~4 of
Ref.~\cite{FHSSW11}. We can then take the convex hull of the achievable rate
regions for these different strategies to get an achievable rate region for a
min-entropy quantum simultaneous decoder (the RHS\ of
Figure~\ref{fig:strong-int} displays an interesting example of such a \textquotedblleft
min-entropy\textquotedblright\ region). If Conjecture~2 of Ref.~\cite{FHSSW11}
regarding the existence of a quantum simultaneous decoder were true, then the
rate region in (\ref{eq:strong-rate-1}-\ref{eq:strong-rate-3})
would be achievable under
under the conditions of (\ref{eq:SI-1}-\ref{eq:SI-2}) (with Holevo
information rates replacing Shannon rates).
Figure~\ref{fig:strong-int}\ displays the different capacity and achievable
rate regions when a free-space interference channel exhibits \textquotedblleft
strong\textquotedblright\ interference.

\section{Han-Kobayashi Rate Regions}

The Han-Kobayashi region is the best known achievable rate region for the
classical interference channel~\cite{HK81}. The coding strategy to achieve
this region is for each receiver to decode partially the other sender's
message while fully decoding the partner sender's message. With this strategy,
the four parties can choose to take advantage of channel interference while
achieving the task of paired sender-receiver communication.

A compact description of the Han-Kobayashi region comes from its reduction
with a Fourier-Motzkin elimination algorithm~\cite{CMGE08}. It is the convex closure of all positive
rate pairs $\left(  R_{1},R_{2}\right)  $ satisfying the following
inequalities and the inequalities obtained from the ones below by swapping the
indices 1 and 2:%
\begin{align}
R_{1} &  \leq I\left(  U_{1}W_{1};B_{1}|W_{2}\right)  ,\label{eq:HK-1}\\
R_{1} &  \leq I\left(  U_{1};B_{1}|W_{1}W_{2}\right)  +I\left(  W_{1}%
;B_{2}|U_{2}W_{2}\right)  ,\\
R_{1}+R_{2} &  \leq I\left(  U_{1};B_{1}|W_{1}W_{2}\right)  +I\left(
U_{2}W_{2}W_{1};B_{2}\right)  ,\\
R_{1}+R_{2} &  \leq I\left(  U_{1}W_{2};B_{1}|W_{1}\right)  +I\left(
U_{2}W_{1};B_{2}|W_{2}\right)  ,\\
2R_{1}+R_{2} &  \leq I\left(  U_{1};B_{1}|W_{1}W_{2}\right)  +I\left(
U_{1}W_{1}W_{2};B_{1}\right)  \nonumber\\
&  \ \ \ \ \ \ \ \ \ +I\left(  U_{2}W_{1};B_{2}|W_{2}\right)  .\label{eq:HK-6}%
\end{align}
In the above, $U_{m}$ is the \textquotedblleft personal\textquotedblright%
\ random variable of Sender~$m$, and $W_{m}$ is her \textquotedblleft
common\textquotedblright\ random variable.

The Han-Kobayashi coding strategy readily translates into a strategy for
coherent-state encoding and coherent detection. Sender~$m$
shares the total photon number $N_{S_{m}}$ between her personal message and
her common message. Let $\lambda_{m}$ be the fraction of signal power that
Sender~$m$ devotes to her personal message, and let $\bar{\lambda}_{m}$ denote
the other fraction of signal power that Sender$~m$ devotes to her common
message. The inequalities above become the following ones for the case of
coherent-state encoding along with coherent detection:%
\begin{gather*}
R_{1}\leq\gamma\left(  \frac{\eta_{11}N_{S_{1}}}{\eta_{21}\lambda_{2}N_{S_{2}%
}+N_{1}}\right)
,\ \ \ \ \ \ \ \ \ \ \ \ \ \ \ \ \ \ \ \ \ \ \ \ \ \ \ \ \ \ \ \ \ \ \\
R_{1}\leq\gamma\left(  \frac{\eta_{11}\lambda_{1}N_{S_{1}}}{\eta_{21}%
\lambda_{2}N_{S_{2}}+N_{1}}\right)  +\gamma\left(  \frac{\eta_{12}\bar
{\lambda}_{1}N_{S_{1}}}{\eta_{12}\lambda_{1}N_{S_{1}}+N_{2}}\right)  ,
\end{gather*}%
\begin{gather*}
R_{1}+R_{2}\leq\gamma\left(  \frac{\eta_{11}\lambda_{1}N_{S_{1}}}{\eta
_{21}\lambda_{2}N_{S_{2}}+N_{1}}\right)
+\ \ \ \ \ \ \ \ \ \ \ \ \ \ \ \ \ \ \ \ \ \ \ \ \ \ \ \ \ \ \ \ \ \ \ \ \ \ \ \ \ \ \ \ \\
\gamma\left(  \frac{\eta_{12}\bar{\lambda}_{1}N_{S_{1}}+\eta_{22}N_{S_{2}}%
}{\eta_{12}\lambda_{1}N_{S_{1}}+N_{2}}\right)  ,\\
R_{1}+R_{2}\leq\gamma\left(  \frac{\eta_{11}\lambda_{1}N_{S_{1}}+\eta_{21}%
\bar{\lambda}_{2}N_{S_{2}}}{\eta_{21}\lambda_{2}N_{S_{2}}+N_{1}}\right)
+\ \ \ \ \ \ \ \ \ \ \ \ \ \ \ \ \ \ \ \ \ \ \ \ \ \ \ \ \ \ \ \ \ \ \\
\gamma\left(  \frac{\eta_{22}\lambda_{2}N_{S_{2}}+\eta_{12}\bar{\lambda}%
_{1}N_{S_{1}}}{\eta_{12}\lambda_{1}N_{S_{1}}+N_{2}}\right)  ,\\
2R_{1}+R_{2}\leq\gamma\left(  \frac{\eta_{11}\lambda_{1}N_{S_{1}}}{\eta
_{21}\lambda_{2}N_{S_{2}}+N_{1}}\right)
+\ \ \ \ \ \ \ \ \ \ \ \ \ \ \ \ \ \ \ \ \ \ \ \ \ \ \ \ \ \ \ \ \ \ \ \ \ \ \ \ \ \ \ \ \ \ \ \\
\gamma\left(  \frac{\eta_{21}\bar{\lambda}_{2}N_{S_{2}}+\eta_{11}N_{S_{1}}%
}{\eta_{21}\lambda_{2}N_{S_{2}}+N_{1}}\right)  +\gamma\left(  \frac{\eta
_{22}\lambda_{2}N_{S_{2}}+\eta_{12}\bar{\lambda}_{1}N_{S_{1}}}{\eta
_{12}\lambda_{1}N_{S_{1}}+N_{2}}\right),
\end{gather*}
where
\[
\gamma\left(  x\right)  =\ln\left(  1+x\right) / 2^i ,\ \ \ \ N_{m}%
=\left[  2^i\bar{\eta}_{m}N_{B_{m}}+1\right] / 4^i ,
\]
$i=1$ for homodyne detection, $i=0$ for heterodyne detection, and
$m\in\left\{  1,2\right\}  $.

We also conjecture a Han-Kobayashi achievable rate region if the senders
employ coherent-state encodings and the receivers exploit joint-detection
receivers (this again follows from Conjecture~2 of Ref.~\cite{FHSSW11}%
\ regarding the existence of a quantum simultaneous decoder). The inequalities
for the region are similar to those in (\ref{eq:HK-1}-\ref{eq:HK-6}) and the
additional \textquotedblleft swapped\textquotedblright\ inequalities, with the
exception that Holevo informations replace mutual informations.

\begin{figure}[ptb]
\begin{center}
\includegraphics[
width=3.5in
]{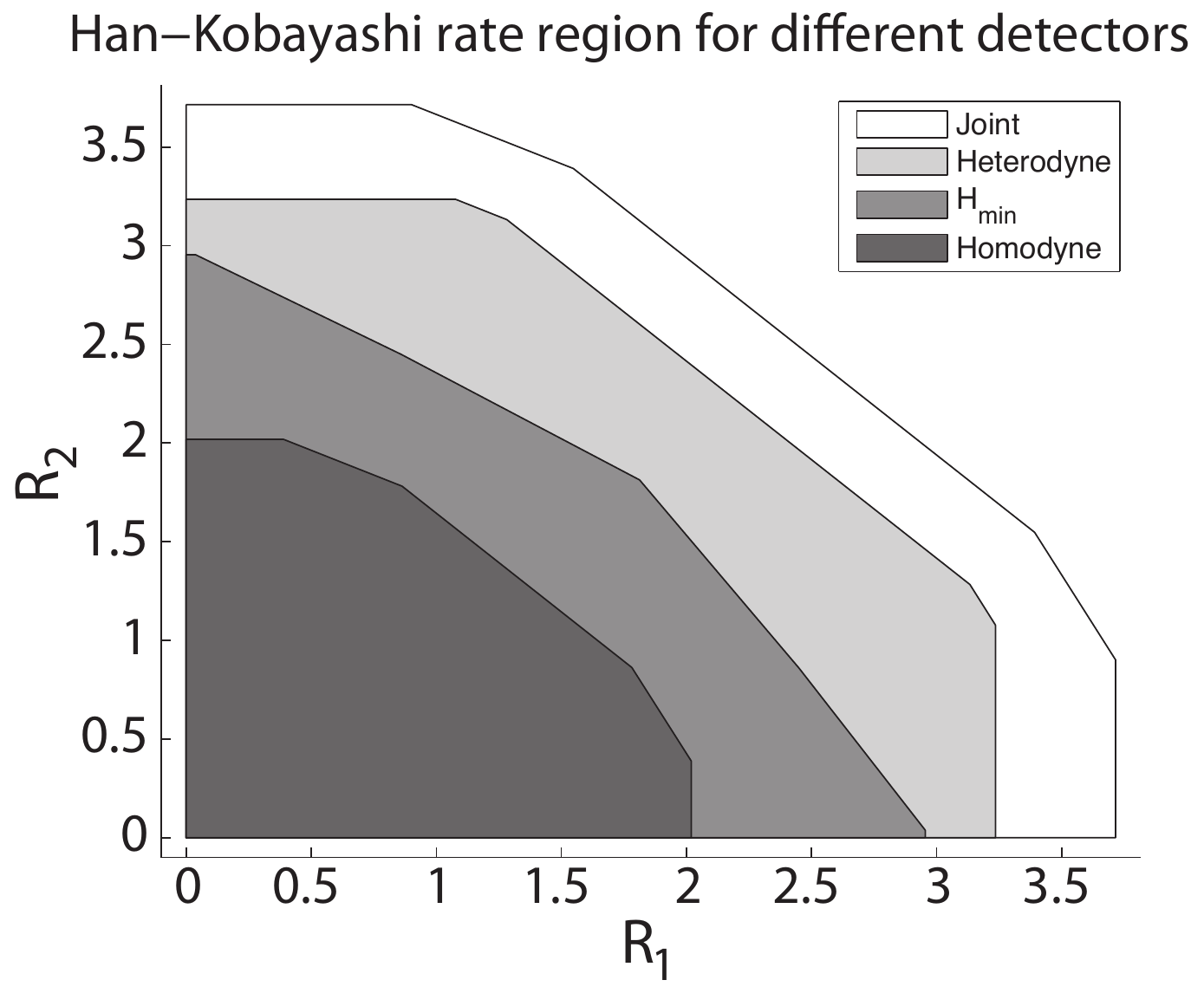}
\end{center}
\caption{The free-space channel no longer exhibits ``very strong'' or ``strong''
interference for coherent detection because the parameters are $N_{S_{1}}=N_{S_{2}%
}=100$, $N_{B_{1}}=N_{B_{2}}=1$, $\eta_{11}=\eta_{22}=0.8$, and $\eta
_{21}=\eta_{12}=0.1$. The figure depicts the achievable rate regions
by employing a Han-Kobayashi coding strategy for homodyne and heterodyne detection.
It also depicts what we can achieve by employing 49 variations of the min-entropy
decoding strategy from Ref.~\cite{FHSSW11} and taking the convex hull of these achievable rate regions.
Finally, it displays the conjectured region if a joint detection decoder
were to exist (see Conjecture~2 of Ref.~\cite{FHSSW11}). All of these regions are with respect to
a 10\%-personal, 90\%-common Han-Kobayashi power split.
%
}%
%
\label{fig:h-k}%
\vspace{-3mm}
\end{figure}

We can obtain an achievable rate region by exploiting the quantum simultaneous
decoder from Theorem~4 of Ref.~\cite{FHSSW11}\ that gives rates which are
a difference of a min-entropy and a von Neumann entropy. The region's characterization
 is in terms of the Han-Kobayashi (HK) characterization with
14 inequalities~\cite{HK81}, corresponding to two different multiple access
channels (MACs) induced to each receiver by the HK coding strategy
(seven~inequalities for each MAC). There are 49 variations
of these min-entropy decoders---one of the information rates
in the seven inequalities for each
MAC is von Neumann and the other six are min-entropy rates.
Then taking the convex hull of these 49 different achievable rate regions gives
an achievable rate region for a min-entropy decoding strategy.
Figure~\ref{fig:h-k} plots the regions achievable with coherent detection,
the min-entropy decoder, and the conjectured joint detector for a particular HK power split.

\vspace{-1mm}

\section{Conclusion}

\vspace{-1mm}

The semiclassical models for free-space optical communication are not
sufficient to understand the ultimate limits on reliable communication rates,
for both point-to-point and multiple-sender-receiver channels. We presented a
quantum-mechanical model for the free-space optical interference channel and determined
achievable rate regions using both structured and unstructured receivers.
Interestingly, the min-entropy decoder from
Ref.~\cite{FHSSW11}\ can achieve rates that are unachievable by
both homodyne and heterodyne detection when the channel exhibits
\textquotedblleft strong interference.\textquotedblright\  Finally, we
determined the Han-Kobayashi inner bound for homodyne and heterodyne
detection, and we conjectured a rate region of this form if a quantum
simultaneous decoder were to exist.

Several open problems remain for this line of inquiry. Perhaps the biggest
open question is to prove Conjecture~2 from Ref.~\cite{FHSSW11}\ concerning
the existence of a quantum simultaneous decoder for a general quantum
interference channel. Also, we do not know if a coherent-state encoding is in
fact optimal for the free-space interference channel---it might be that
squeezed state transmitters could achieve higher communication rates as in
Ref.~\cite{Y05}. One could also evaluate the ergodic and outage capacity
regions based on the statistics of $\eta_{ij}$, which could be derived from
the spatial coherence functions of the stochastic mode patterns under
atmospheric turbulence.

We ackowledge useful discussions with K.~Br\'{a}dler, O.~Fawzi,
P.~Hayden, P.~Sen, and B.~Yen. S.~Guha acknowledges the
DARPA Information in a Photon program, contract \#HR0011-10-C-0159.
M.~M.~Wilde\ acknowledges the MDEIE (Qu\'{e}bec) PSR-SIIRI international collaboration grant.
I.~Savov acknowledges support from FQRNT and NSERC.

\bibliographystyle{IEEEtran}
\bibliography{Ref}

\end{document}